\documentclass[12pt]{article}
\usepackage[dvips]{graphicx}
\usepackage{latexsym}
\usepackage{amssymb}

\newcommand{\CP}{\mathbb{CP}}

\newcommand{\R}{\mathbb{R}}

\def\p{\partial}

\def\bb{{\tilde \beta}}
\newcommand{\bea}{\begin{eqnarray}}
\newcommand{\eea}{\end{eqnarray}}

\newcommand{\koniec}{\begin{flushright}  $\Box $ \end{flushright}}
\def\be{\begin{equation}}
\def\ee{\end{equation}}

\def\p{\partial}

\newtheorem{theo}{Theorem}[section]

\newcounter{mnotecount}[section]

\renewcommand{\themnotecount}{\thesection.\arabic{mnotecount}}

\def\bbe{{\bf{e}}}
\def\hn {\hat{\nabla}}
\font\mybb=msbm10 at 11pt

\def\bb#1{\hbox{\mybb#1}}

\def\bR {\bb{R}}

\def\bC {\bb{C}}
\newcommand{\nn}{\nonumber \\}

\def\hp {\hat{\partial}}

\def\bcc#1{\buildrel \circ \over #1}

\def\bcn {{\bcc \nabla}}

\def\tn {{\hat{\nabla}}}

\def\hn {{\hat{\nabla}}}

\def\dh {{\dot{h}}}
\def\ddh {{\ddot{h}}}
\def\de {{\dot{\bf{e}}}}

\def\rsq{{\cal{O}}(r^2)}

\def\appendix#1{\addtocounter{section}{1}\setcounter{equation}{0}
\renewcommand{\thesection}{\Alph{section}}
\section*{Appendix \thesection\protect\indent \parbox[t]{11.15cm}{#1}}
\addcontentsline{toc}{section}{Appendix \thesection\ \ \ #1}}

\newcommand{\mnote}[1]
{\protect{\stepcounter{mnotecount}}$^{\mbox{\footnotesize
$
\bullet$\themnotecount}}$ \marginpar{
\raggedright\tiny\em
$\!\!\!\!\!\!\,\bullet$\themnotecount: #1} }

\begin{document}
\pagestyle{plain}
\title{\vskip -60pt
\begin{flushright}
{\normalsize DAMTP-2016-71},  {\normalsize DMUS--MP--16/21}\\
\end{flushright}
\vskip 8pt
{\bf Einstein--Weyl Spaces and Near-Horizon Geometry}
\vskip 3pt}
\author{Maciej\ Dunajski
\thanks{email {\tt m.dunajski@damtp.cam.ac.uk}}\\
{\sl Department of Applied Mathematics and Theoretical Physics} \\
{\sl University of Cambridge} \\
{\sl Wilberforce Road, Cambridge CB3 0WA, UK.} \\[8pt]
Jan \ Gutowski \thanks{email {\tt j.gutowski@surrey.ac.uk}}\\
{\sl
Department of Mathematics, University of Surrey }\\
{\sl  Guildford, GU2 7XH, UK.}\\[8pt]
Wafic \ Sabra \thanks{email {\tt ws00@aub.edu.lb}}
\\[3pt]
{\sl  Centre for Advanced Mathematical Sciences and Physics Department}\\[1pt]
{\sl American University of Beirut, Beirut, Lebanon.}
}
\date{October 27, 2016}
\maketitle
\thispagestyle{empty}
\begin{abstract}
We show that a class of solutions of minimal supergravity in five dimensions
is given by lifts of three--dimensional Einstein--Weyl structures of hyper-CR type.
We characterise this class as most general near--horizon limits of supersymmetric
solutions to the five--dimensional theory. In particular we deduce that a
compact spatial section of a horizon
can only be a Berger sphere, a product metric on $S^1\times S^2$ or a flat three-torus.

We then consider the problem of reconstructing all supersymmetric solutions from a given near--horizon geometry.
By exploiting the ellipticity of the linearised field equations we demonstrate that the moduli space
of transverse infinitesimal deformations of a near--horizon geometry is finite--dimensional.

\end{abstract}
\newpage

\maketitle
\section{Introduction}

In dimension four the topology of a black--hole horizon is necessarily spherical, but in dimensions five and higher this restriction needs not to hold. The local differential geometric structure of horizons in higher dimensional gravity theories is also rich. The aim of this paper is to demonstrate that in case of supersymmetric solutions of
minimal supergravity  in dimension five, the horizon geometry is that of a three-dimensional Riemannian Einstein--Weyl structure of hyper-CR type. We shall also show that the moduli space of linearised transverse deformations of near-horizon geometries with compact spatial sections of horizons
is finite--dimensional.

In the first part of the paper we shall demonstrate that a class  of solutions of minimal supergravity in five dimensions
is given by lifts of three--dimensional Einstein--Weyl structures of hyper-CR type.
We characterise this class as most general near--horizon limits of supersymmetric
solutions to the five--dimensional theory. In particular we deduce that a
compact spatial section of a horizon
can only be a Berger sphere, a product metric on $S^1\times S^2$ or a flat three-torus.

Space times containing Killing horizons with vanishing surface gravity admit a
limiting procedure leading to a near horizon geometry \cite{harvey, LP, MI08,
KL}.  In the second part of the paper we consider the problem of reconstructing all supersymmetric solutions with a given near--horizon geometry.  By exploiting the ellipticity of the linearised field equations we demonstrate that the moduli space of infinitesimal transverse deformations of a near--horizon geometry is finite--dimensional if the spatial section of the horizon is compact.

In the next two sections we shall review the  $D=5$ and $N=2$ supergravity with its near horizon limit,
and the hyper--CR Einstein Weyl equations respectively. In section \ref{sec_result} we shall
establish (Theorem \ref{main_theo_part1}) a correspondence between 3D hyper-CR Einstein--Weyl structures and near-horizon limits of supersymmetric solutions to $D=5$, $N=2$ supergravity, and deduce the allowed topologies of horizons in this case. We shall also construct an explicit local fibration
of the 5D metric over a hyper--K\"ahler four--manifold with a homothetic Killing vector field.

In section \ref{section5} we shall prepare the ground for the moduli space calculation, and derive the Bianchi identities resulting from supersymmetry. In section \ref{section6} (Theorem \ref{theorem2}) we shall prove that, if the  spatial horizon 3-surface $\Sigma\subset M$ is compact, then the moduli space of infinitesimal
transverse
deformations of near-horizon geometry is finite--dimensional.

The long formulae involving the Ricci tensor in the bulk, and the analysis of the gravitino equation
have been relegated to Appendices.

\section{5D Minimal supergravity and its near-horizon limit}

Let $M$ be a five--dimensional manifold with pseudo--Riemannian metric $g$ of signature
$(4, 1)$ and a Maxwell potential $A$. The five--dimensional action for the Einstein--Maxwell theory with a Chern--Simons term
is
\be
\label{lagrangian}
S=\int_M {\mathcal R}\;\mbox{vol}_M-\frac{3}{2}H\wedge \ast_5 H-
H\wedge H\wedge A,
\ee
where ${\mathcal R}$ is the Ricci scalar of $g$, $H=dA$ is the $U(1)$ Maxwell field, $\mbox{vol}_M$ is the volume element induced by $g$
and $\ast_5:\Lambda^\alpha(M)\rightarrow \Lambda^{5-\alpha}(M)$ is the associated Hodge endomorphism.
The resulting Einstein--Maxwell--Chern--Simons equations coincide with the bosonic sector
of minimal five--dimensional supergravity. These equations are
\begin{eqnarray}
\label{EMCS}
&&dH=0, \quad d\ast_5 H+ H\wedge H=0,\nonumber\\
&&{\mathcal R}_{\alpha\beta}-\frac{3}{2}H_{\alpha\gamma}{H_\beta}^{\gamma}+\frac{1}{4}g_{\alpha\beta}H^2=0.
\end{eqnarray}
In \cite{QMW} it was shown that all supersymmetric solutions  of this theory
admit local fibrations over hyper--K\"ahler four manifolds, i. e.
locally  there exists a function $u:M\rightarrow\R$ such that
\be
\label{HKfib}
g=-f^2(du-\Theta)^2+f^{-1}g^{HK},
\ee
where $g^{HK}$ is a $u$--independent hyper-K\"ahler metric on some four--manifold $X$,
and $(\Theta, f)$ are respectively a one--form and a function on $X$ which do not depend on $u$.

\subsection{Near-horizon limit}
Let $(u, r, y^i)$, where $i=1, 2, 3$, be Gaussian null coordinates \cite{MI08}
defined in the neighbourhood of a Killing horizon $g(V, V)=0$ where $V=\p/\p u$
is a stationary Killing vector. In these coordinates the horizon is given by $r=0$, and
$y^i$ are local coordinates on three--dimensional Riemannian manifold $\Sigma$ which is
the spatial section of the horizon. The metric and the Maxwell potential are given by
\be
\label{gaussian}
g=2du\Big(dr+rh-\frac{1}{2}r^2\Delta du\Big)+\gamma, \quad
A=r\Phi du+B,
\ee
where
\[
\gamma=\gamma_{ij}(r, y^k)dy^idy^j,\quad h=h_i(r, y^k)dy^i, \quad B=B_i(r, y^k) dy^i,
\quad \Delta=\Delta(r, y^k), \quad \Phi=\Phi(r, y^k)
\]
are all real--analytic in $r$.
The near horizon limit \cite{harvey, LP, MI08, KL} arises by replacing
\[
u\longrightarrow u/\epsilon, \quad r\longrightarrow r\epsilon,
\]
and taking the limit $\epsilon\rightarrow 0$. The resulting metric and
Maxwell potential
on the neighbourhood of the horizon in  $M$ are given by (\ref{gaussian}), where
now $\gamma$ is a Riemannian metric, $(h, B)$ are one--forms and $(\Phi, \Delta)$ are functions
on $\Sigma$ which depend on the local coordinates $y^k$, but not on $(r, u)$.
In the $\epsilon\rightarrow 0$ limit, the Euler--Lagrange equations of the functional (\ref{lagrangian}) yield a set of equations on the three--dimensional data on $\Sigma$.
The Maxwell--Chern--Simons equations are
\be
\label{gauge_eq}
d\ast_3 dB+\ast_3(d\Phi-\Phi h)-h\wedge \ast_3 dB-2\Phi dB=0,
\ee
and the non--trivial components of the Einstein equations are
\be
\label{einstein_1}
\frac{1}{2} \nabla^ih_i-\frac{1}{2}h^ih_i+\frac{1}{4}dB_{ij}dB^{ij}+\Phi^2-\Delta=0,
\ee
and
\be
\label{einstein_2}
R_{ij}+\nabla_{(i}h_{j)}-\frac{1}{2}h_{i} h_{j}-\frac{3}{2}dB_{ik}{dB_j}^k+\gamma_{ij}
\Big(\frac{1}{4}dB_{kl}dB^{kl}-\frac{1}{2}\Phi^2\Big)=0.
\ee
Here $\nabla$ is the Levi--Civita connection of the metric $\gamma=\gamma_{ij}(y)dy^idy^j$
and $\ast_3:\Lambda^i(\Sigma)\rightarrow \Lambda^{3-i}(\Sigma)$ is the Hodge operator on $\Sigma$.
The necessary conditions for the near--horizon geometry to be supersymmetric
are \cite{jan}
\be
h+\ast_3 dB=0, \quad \Delta=\Phi^2.
\label{jan}
\ee
\section{Hyper--CR Einstein--Weyl geometry}
A Riemannian Weyl structure on a three--dimensional manifold $\Sigma$ consists of a positive--definite conformal structure
$[\gamma]=\{c\gamma, c:\Sigma\rightarrow \R^+\}$, and a torsion--free connection $D$ which is compatible with $[\gamma]$ is the sense that
\[
D_i\gamma_{jk}=2h_i  \gamma_{jk},
\]
for some one--form $h$ on $\Sigma$. This compatibility condition is invariant under the transformation
\be
\label{ew_transform}
\gamma\rightarrow e^{2\Omega} \gamma, \quad h\rightarrow h+d\Omega,
\ee
where
$\Omega$ is a function on $\Sigma$. A choice of the conformal factor $\Omega$ such that
\be
\label{GG}
\nabla^i h_i=0
\ee
is called the Gauduchon gauge.

 A Weyl structure is said to be Einstein--Weyl \cite{Cartan, hitchin} if the symmetrised Ricci tensor of $D$
is proportional to some metric $\gamma\in [\gamma]$. This conformally invariant
condition can be formulated directly as a set of non--linear PDEs on the pair
$(\gamma, h)$:
\be
\label{ew_eq}
R_{ij}+\nabla_{(i}h_{j)}+h_ih_j-\frac{1}{3}
\Big(R+\nabla^k h_k+h^k h_k\Big) \gamma_{ij}=0,
\ee
where $\nabla$, $R_{ij}$, and $R$ are respectively the Levi--Civita connection, the Ricci tensor and the Ricci scalar of $\gamma$.

A tensor object $T$ which transforms like $T\rightarrow \exp{(m\Omega}) T$ when $\gamma\rightarrow \exp({2\Omega})\gamma$
is said to be conformally invariant with weight $m$. The Ricci scalar $W$ and the Ricci tensor $W_{ij}$ of
the Weyl connection have weights $-2$ and $0$ respectively. The Ricci scalar is given by
\be
\label{weyl_scalar}
W=R+4\nabla^i h_i-2 h^ih_i.
\ee
An Einstein--Weyl space is called\footnote{This class of Einstein--Weyl spaces
has originally been called `special' in \cite{GT}, and it has also been referred to as `Gauduchon--Tod'.
The current terminology  (see e. g. \cite{DT01, CP, Dhydro, DK}) reflects the fact that the hyper--CR EW spaces
arise as symmetry reductions of four--dimensional hyper--complex conformal structures by tri--holomorphic isometry.
The three--dimensional quotients admit a sphere of Cauchy--Riemann structures.}
hyper-CR if \cite{GT} there exists a scalar function $\Phi$ of weight $-1$ which, together with the
EW one--form $h$, satisfies the monopole equation
\be
\label{CR_monopole}
\ast_3(d \Phi+ h\Phi)=dh
\ee
together with an algebraic constraint
\be
\label{alg_cons}
W=\frac{3}{2}\Phi^2.
\ee
The hyper--CR Einstein--Weyl spaces can be equivalently characterised by
the existence of a holomorphic fibration of the associated mini--twistor spaces
over $\CP^1$, or by existence of two--parameter family of shear--free, divergence--free
geodesic congruences \cite{CP}. The only compact examples are the Berger sphere, $S^1\times S^2$ or $T^3$ with the
flat EW structure.

In the real--analytic category,
a hyper--CR EW structure locally depends on two arbitrary functions of two variables.
This can be seen by reformulating the hyper--CR condition in terms of a single
second order PDE for one function of three variables \cite{DT01}:
\be
\label{local1}
\gamma= dzd\bar{z}+\frac{1}{16}(F d v-i(F_zdz-F_{\bar{z}} d\bar{z})+dF_v)^2,
\quad h=\frac{(F_z+iF_{vz})dz+(F_{\bar{z}}-iF_{v\bar{z}})d\bar{z}}{F+F_{vv}},
\ee
where $F=F(z, \bar{z}, v)$ satisfies
\be
\label{local2}
F_{z\bar{z}}(F+F_{vv})-(F_z+iF_{vz})(F_{\bar{z}}-iF_{v\bar{z}})=4.
\ee
The Cauchy--Kowalewskaya Theorem implies that the arbitrary data
\[
F(z, \bar{z}, v=0), \quad\mbox{and}\quad F_v(z, \bar{z}, v=0)
\]
specifies the solution uniquely.

\section{From 3D Einstein--Weyl to 5D minimal supergravity}
\label{sec_result}
We shall now show that hyper--CR Einstein--Weyl spaces
give rise to solutions of minimal $D=5$ SUGRA, and characterise the solutions which are obtained
from this procedure. Roughly speaking, any
hyper--CR Einstein--Weyl structure lifts to a solution
of the five-dimensional theory, provided that the Gauduchon gauge is chosen. This reflects the fact that (unlike
the EW equations) the 5D SUGRA equations are not conformally invariant. To overcome this we will need to introduce
and solve a linear PDE ({\ref{GG_pde}) on the EW background to achieve the right gauge fixing.
\begin{theo}
\label{main_theo_part1}
Let the metric $\gamma$ and the one--form $h$ on a three--dimensional manifold $\Sigma$
solve the hyper--CR Einstein--Weyl equations, and let $W$ be the Ricci scalar
of the Weyl connection of $(\gamma, h)$
given by (\ref{weyl_scalar}).
Let $\Omega$ be a function on $\Sigma$ which satisfies
the  linear PDE
\be
\label{GG_pde}
d\ast_3  \big(d e^\Omega \big) +d\ast_3 \big( e^\Omega h \big) =0 \ .
\ee
Then
\be
\label{final}
g=e^{2\Omega}( 2du(dr+rh-\frac{1}{3}r^2 Wdu)+\gamma  +6rdud\Omega), \quad A=
\sqrt{\frac{2}{3}}e^\Omega r\sqrt{W}du
+\alpha
\ee
is a solution to the 5D Einstein--Maxwell--Chern--Simons
supergravity (\ref{EMCS}). Here $\alpha$ is a one--form on $\Sigma$ such that
\be
\label{maxwell}
d\alpha=-e^{\Omega}\ast_3(h+d\Omega) \ .
\ee

  All near--horizon geometries for 5D SUSY back holes/rings/strings
(\ref{gaussian}) are locally of the form (\ref{final}). Moreover if the
three--manifold corresponding to the spatial sections of the horizon
is compact, then $\gamma$ is a metric on the Berger sphere,
a product metric on $S^1\times S^2$ or a flat metric on $T^3$.
\end{theo}
{\bf Proof.}
Consider the field equations (\ref{gauge_eq}, \ref{einstein_1}, \ref{einstein_2})
and additionally assume that the SUSY constrains (\ref{jan}) hold.
Thus $(dB)_{jk}=-\epsilon_{ijk}h_i$ and
\[
dB_{im}{dB_j}^m=(\gamma_{ij}|h|^2-h_ih_j) \ .
\]
The Maxwell--Chern--Simons condition (\ref{gauge_eq}) now reduces to the
monopole equation (\ref{CR_monopole}). We regard
$\Phi$ as a weighted scalar with conformal weight $(-1)$
on the three--manifold $\Sigma$.
The $(ur)$ component (\ref{einstein_1}) of the Einstein equation becomes
(\ref{GG}). Thus the five--dimensional field equations force the Gauduchon
gauge on the Weyl geometry. Finally the $(ij)$ components
(\ref{einstein_2}) of the Einstein equations yield
\be
\label{in_the_proof}
R_{ij}+\nabla_{(i}h_{j)}+h_ih_j=
\Big(\frac{1}{2}\Phi^2+h^k h_k\Big) \gamma_{ij}.
\ee
Taking a trace of this condition and using the Gauduchon gauge (\ref{GG})
gives a constraint on the Ricci scalar of $\gamma$
\[
R=\frac{1}{2}(3\Phi^2+4|h|^2).
\]
Thus $R\geq 0$ with the equality iff both $\gamma$ and $D$ on $\Sigma$ are flat.
We can now use the expression (\ref{weyl_scalar})
for the Ricci scalar  of the Weyl connection
to re-express this constraint as (\ref{alg_cons}).
Now equations (\ref{in_the_proof}) are equivalent to the Einstein--Weyl
equations (\ref{ew_eq}) in the Gauduchon gauge, subject to two constraints
(\ref{alg_cons}) and (\ref{CR_monopole}). These two constrains are (as we have explained in the previous section) the defining property of the hyper--CR
Einstein--Weyl conditions.

To recover the form of the five--dimensional solution from the hyper--CR
EW geometry we must make sure that the latter is given in the Gauduchon gauge.
This appears to break the conformal invariance of our procedure, but it
was to be expected as the five--dimensional theory is not conformally invariant.
Assume that a hyper--CR EW structure $(\hat{\gamma}, \hat{h})$ is given in
the Gauduchon gauge. Substituting the Einstein--Weyl data into (\ref{gaussian})
we find that the construction explained so far gives
the lift to five dimensions of the form
\be
\label{ghat}
{g}=2d\hat{u}(d\hat{r}+\hat{r}\hat{h}-\frac{1}{3}\hat{r}^2 \hat{W}d\hat{u})+\hat{\gamma},
\ee
where $(\hat{r}, \hat{u})$ are some local Gaussian coordinates in the neighbourhood of the horizon.
Now consider a hyper--CR EW structure $(\gamma, h)$
in an arbitrary gauge. To put it in a Gauduchon gauge
we need to find a function $\Omega$ on $\Sigma$ such that
(\ref{GG_pde}) holds. The solution of this equation always exists locally on
$\Sigma$,  and it follows from the work of Tod \cite{tod_cpt} that it also exists
globally on compact EW manifolds. Thus the EW structure
\be
\label{gauge_proof}
\hat{h}=h+d\Omega, \quad \hat{\gamma}=e^{2\Omega}\gamma
\ee
is in the Gauduchon gauge. The constraints
(\ref{alg_cons}) and (\ref{CR_monopole}) are preserved
under the conformal rescaling if $\Phi$ has conformal weight (-1).
Therefore we substitute (\ref{gauge_proof}) together with
\[\hat{\Phi}=e^{-\Omega}\Phi\] into (\ref{ghat}). We also make coordinate
changes \[(\hat{r}, \hat{u})=(e^{2\Omega}r, u).\] This yields
(\ref{final}). We now consider the gauge potential.
We substitute the expression (\ref{jan}) for $dB$ into $H=dA$, where $A$ is
 given by (\ref{gaussian}). Using the equation (\ref{CR_monopole})
and tracking down the effect of conformal resealing needed
to enforce the Gauduchon gauge yields ${A}$ in (\ref{final}).
\koniec
{\bf Remarks}
\begin{itemize}
\item
As a spin-off from this analysis we have established the
transformation rule for the five--dimensional structures $(g, A)$ under the conformal rescalings  (\ref{ew_transform}) of the underlying EW geometry. We see that
the metric $g$ does not transform by a simple scaling but also picks up an inhomogeneous
term
\[
g\longrightarrow e^{2\Omega}(g+6 r du d\Omega) .
\] This additional term of course
vanishes on the horizon.
\item Three--dimensional Einstein--Weyl equations are integrable by twistor transform \cite{hitchin},
and can be regarded as a master dispersionless integrable system  in 3 and 2+1 dimensions - all other known dispersionless systems arise as special cases.
It is remarkable that in case of super--symmetric solutions the non--integrable equations
of 5D minimal supergravity reduce to integrable Einstein--Weyl structures. This phenomenon has been observed in other supergravity theories \cite{GGS, DGST1, DGST2, messenortin} supporting the
evidence that that the supersymmetric sectors of non--integrable classical field theories can be described by integrable models.
\end{itemize}
\subsection{Fibration over a hyper--K\"ahler manifold}
Finally we shall show how to put the metric ${g}$  (\ref{final}) in the form of the fibration
(\ref{HKfib}) over a hyper--K\"ahler manifold. Comparing the expressions
(\ref{final}) and (\ref{HKfib}), and completing the square  yields
\[
f^2=\frac{2}{3}e^{2\Omega} r^2W.
\]
We now define a new coordinate $\rho$ by
$r=\exp{(\rho-3\Omega)}$ and find
\be
\label{HK_tri}
g^{HK}=e^{\rho}(\Phi \gamma+\Phi^{-1}(d\rho+h)^2), \quad
f=\Phi e^{\rho-2\Omega}, \quad \Theta=\Phi^{-2}e^{3\Omega-\rho}(d\rho+h),
\ee
where $\Phi^2=2W/3$ and $W$ is the scalar curvature of the Weyl connection
of $(\gamma, h)$ given by (\ref{weyl_scalar}). Moreover, it follows from
the work of \cite{GT} that any hyper--Kahler metric with a tri--holomorphic
homothety\footnote{Recall that a tri--holomorphic homothety
is a conformal Killing vector $K$ which preserves the sphere of complex
structures, i. e
\[
{\mathcal L}_K(g^{HK})=\eta \;g^{HK}, \quad {\mathcal L}_K I_i=0,
\]
where $\eta$ is a non--zero constant, and $I_i, i=1, 2, 3$ are the complex
structures satisfying the quaternionic algebra.}
is of the form (\ref{HK_tri}) for some Hyper--CR EW structure
$(\gamma, h)$.
In our coordinate system the homothety is generated by
$\p/\p \rho$, and the horizon in five dimensions corresponds to
$\rho=-\infty$.

\section{Extension into the Bulk}
\label{section5}

In this section we briefly summarise some of the conditions imposed on the gauge field
strength and the geometry by supersymmetry which we shall use in the
moduli calculation in section \ref{section6}. The  gravitino Killing spinor equation is given by:
\begin{equation}
\left[ \nabla_{\alpha}-{\frac{i}{8}}\Gamma_{\alpha}H_{\nu_{1}\nu_{2}}\Gamma
^{\nu_{1}\nu_{2}}+{\frac{3i}{4}}H_{\alpha}{}^{\nu}\Gamma_{\nu} \right] \epsilon =0,  \label{grav}
\end{equation}
where $\epsilon$ is a Dirac spinor. Here we work with a metric of mostly plus signature $(-,+,+,+,+)$.

We have already introduced the Gaussian null coordinates.
In what follows, it is convenient to adopt the basis $\{ \bbe^+, \bbe^-, \bbe^i : i=1,2,3 \}$,
adapted to the Gaussian null co-ordinate system
in which
\bea
\label{frame}
g = 2 \bbe^+ \bbe^- + \delta_{ij} \bbe^i \bbe^j
\eea
and
\bea
\bbe^+ &=& du
\nonumber \\
\bbe^- &=& dr +rh -{1 \over 2} r^2 \Delta du
\nonumber
\eea and take a
$u$-independent basis of ${{\Sigma}}$ to be given by $\bbe^i= \bbe^i{}_j dy^j$
for $i, j=1,2,3$;
where $\bbe^i{}_j$ depends analytically on $r$, but not on $u$ and $y^j$ are local co-ordinates on ${\Sigma}$. The spin connection associated with the above frame
is computed in Appendix C.

Furthermore, in what follows,  $\ \dot{} \ $ denotes the Lie derivative with respect to ${\partial \over \partial r}$;
$\hat{d}$ denotes the restriction of the exterior derivative to surfaces of constant $r$,
i.e.
\bea
{\hat{d}} \Delta = \partial_i \Delta dy^i,
\qquad
{\hat{d}} h = {1 \over 2} (\partial_i h_j - \partial_j h_i) dy^i \wedge dy^j, \qquad \partial_i = {\partial \over \partial y^i}.
\eea


\subsection{Conditions obtained from Supersymmetry}

Supersymmetry imposes a number of conditions on the gauge field strength,
as well as conditions on the geometry. The explicit analysis of the
Killing spinor equations is given in Appendix C. Here we shall summarise
the results which will be of use in the moduli space analysis to follow.
In particular, the gauge field strength is given by
\bea
H= \eta du \wedge d \big(r \Delta^{1 \over 2} \big)
+{1 \over 3} \bigg(\star_3 Y + (dr+rh) \wedge W \bigg)
\eea
where $W$ is a $r$-dependent 1-form on $\Sigma$, and
\bea
\label{ident1}
Y = -3 h - 3r \dh +2 \eta r \sqrt{\Delta} W.
\eea
The Bianchi identity associated to this expression for $H$ implies
\be
\label{Bianchi1}
\hat{d}( rh\wedge W + \star_3 Y ) = 0
\ee
and
\be
\label{Bianchi2}
\hat{d}W - {\cal{L}}_{\partial \over \partial r}(r h\wedge W + \star_3 Y ) = 0.
\ee
A number of further useful identities obtained from the supersymmetry analysis
are
\bea
\label{ident2}
{\hat{d}} h = r h \wedge \dh
- \eta \star_3 \bigg(\Delta^{1 \over 2} h +2r \Delta^{1 \over 2}
\dh - \eta r \Delta W + {1 \over 2} \Delta^{-{1 \over 2}} {\hat{d}} \Delta
-{1 \over 2} \eta \Delta^{-{1 \over 2}} {\dot{\Delta}} h \bigg)
\eea
\bea
\label{ident3}
{\hat{d}}Y &=& -3 {\cal{L}}_{\partial \over \partial r}
\bigg(r^2 h \wedge \dh
- \eta r \star_3 \big(\Delta^{1 \over 2} h +2r \Delta^{1 \over 2}
\dh - \eta r \Delta W + {1 \over 2} \Delta^{-{1 \over 2}} {\hat{d}} \Delta
-{1 \over 2} r \Delta^{-{1 \over 2}} {\dot{\Delta}} h \big) \bigg)
\nonumber \\
&+& 2 \eta r {\hat{d}} \big(\Delta^{1 \over 2} W \big).
\eea
Also, the gauge field equations reduce to the following:
\be
\label{gauge1}
\hat{d}\star_3W - {\cal{L}}_{\partial \over \partial r}(r h\wedge \star_3W) + {2\over3}W\wedge\star_3 Y  - 3\eta {\cal{L}}_{\partial \over \partial r}(\partial_r(r\Delta^{{1\over2}})\ {\rm dvol}_{{\Sigma}}) = 0 \ .
\ee


\section{Moduli Space Calculation}
\label{section6}

We shall now consider the moduli space of infinitesimal supersymmetric transverse deformations of the near-horizon data, and prove that, for compact $\Sigma$, this is finite-dimensional by establishing that the moduli are constrained by certain elliptic second order differential operators. This analysis
follows that done in \cite{lucietti} for the case of non-supersymmetric vacuum horizons with a cosmological constant, though for the solutions
we consider, there is some modification due to the inclusion of a 2-form
field strength, as well as supersymmetry.

In particular, suppose that we consider the metric written in Gaussian null coordinates
as  (\ref{gaussian})
and Taylor expand the metric data $(\Delta, h, \gamma)$ as
\bea
\Delta &=& \bcc{\Delta}(y)+ r \delta \Delta (y) + \rsq,
\nonumber \\
h &=& \bcc{h} (y) + r \delta h (y) + \rsq,
\nonumber \\
\gamma &=& \bcc{\gamma}(y)+ r \delta \gamma(y) + \rsq
\eea
where $\bcc{\Delta}, \bcc{h}, \bcc{\gamma}$ are the near-horizon metric data,
and the metric moduli are $\delta \Delta, \delta h, \delta \gamma$. There is
some gauge ambiguity in this choice of metric moduli, although the
near horizon data is unique. As noted in \cite{lucietti},
the vector field
\bea
\xi = {1 \over 2} f \bigg(dr+r \bcc{h}-{1 \over 2} r^2 \bcc{\Delta} du \bigg)
-{1 \over 4} r^2 \bigg(\bcc{\Delta} f + {\cal{L}}_{\bcc{h}} f \bigg) du -{1 \over 2} r df
\eea
for an arbitrary smooth function $f$ on $\Sigma$, maps the near-horizon data
$(\delta \Delta, \delta h, \delta \gamma)$ to $(\delta {\tilde{\Delta}},
\delta {\tilde{h}}, \delta {\tilde{\gamma}})$ where

\bea
\label{gtrans}
\delta {\tilde{\gamma}}_{ij} &=& \delta \gamma_{ij} + \bcc{\nabla}_i \bcc{\nabla}_j f
- \bcc{h}_{(i} \bcc{\nabla}_{j)} f
\nonumber \\
\delta {\tilde{h}}_i &=& \delta h_i +{1 \over 2} \bcc{\Delta} {\bcc{\nabla}}_i f
-{1 \over 4} ({\bcc{\nabla}}_i \bcc{h}_j) {\bcc{\nabla}}^j f
-{1 \over 4} \bcc{h}_i \bcc{h}_j {\bcc{\nabla}}^j f
+{1 \over 2} ({\bcc{\nabla}}_j \bcc{h}_i) {\bcc{\nabla}}^j f
+{1 \over 4} \bcc{h}_j {\bcc{\nabla}}_i {\bcc{\nabla}}^j f
\nonumber \\
\delta {\tilde{\Delta}} &=& \delta \Delta +{1 \over 2} {\bcc{\nabla}}^i f
\bigg({\bcc{\nabla}}_i \bcc{\Delta} - \bcc{h}_i {\bcc{\Delta}} \bigg).
\eea

In addition to the metric, we also have the Maxwell 2-form, which we have
shown can be decomposed as
\bea
H= \eta du \wedge d \big(r \Delta^{1 \over 2} \big)
+{1 \over 3} \bigg(\star_3 \bigg(-3h -3r {\dot{h}} +2 \eta r
\sqrt{\Delta} W \bigg) + (dr+rh) \wedge W \bigg)
\eea
where $W$ is a 1-form on $\Sigma$.
The $--$ component of the Einstein equations implies that $W$ is of
the same order as $(\delta \Delta, \delta h, \delta \gamma)$.
We therefore take the transverse moduli to be $(\delta \Delta, \delta h, \delta \gamma, W)$

\begin{theo}
\label{theorem2}
The moduli space of supersymmetric transverse deformations of
supersymmetric near horizon solutions with compact spatial sections of
horizons, corresponding to
the moduli $(\delta \Delta, \delta h, \delta \gamma, W)$,
modulo the gauge transformations of the type ({\ref{gtrans}}), is finite dimensional.
\end{theo}
{\bf Proof.}  To begin, we shall consider the trace of the metric moduli $\delta \gamma_{ij}$,
and prove that by choosing an appropriate gauge transformation ({\ref{gtrans}}), this
modulus satisfies an elliptic PDE which decouples from the remaining transverse moduli.

In particular, the trace transforms as
\bea
\label{trtrans}
\delta \gamma_k{}^k \rightarrow \delta \gamma_k{}^k + {\cal{D}} f
\eea
where
\bea
{\cal{D}} \equiv \bcn^2 - {\bcc h}^i \bcn_i \ ,
\eea
and the adjoint is given by
\bea
{\cal{D}}^\dagger = \bcn^2 + {\bcc h}^i \bcn_i
\eea
because $\bcn_i {\bcc h}^i=0${\footnote{This was established in \cite{harvey}.}}. We decompose $\delta \gamma_k{}^k$
as
\bea
\delta \gamma_k{}^k = \phi + \phi^\perp
\eea
where $\phi \in {\rm Im} {\cal{D}}$, and $\phi^\perp \in \big({\rm Im} {\cal{D}} \big)^\perp$. It follows that
\bea
\phi = {\cal{D}} (\tau)
\eea
for some smooth function $\tau$, and $\phi^\perp \in \big({\rm Im} {\cal{D}} \big)^\perp$ must satisfy
\bea
{\cal{D}}^\dagger \phi^\perp =0 \ ,
\eea
which is an elliptic PDE.
So, choosing $f=-\tau$ in the transformation ({\ref{trtrans}}), with this choice of gauge
\bea
\delta \gamma_k{}^k = \phi^\perp
\eea
and so in this gauge
\bea
\label{trelliptic}
{\cal{D}}^\dagger \delta \gamma_k{}^k=0 \ .
\eea
In fact, this implies that $\delta \gamma_k{}^k$ must be constant. To see this,
note that ({\ref{trelliptic}}) implies that
\bea
\delta \gamma_k{}^k \bcn^2 \delta \gamma_j{}^j + {1 \over 2} {\bcc h}^i \bcn_i
\bigg((\delta \gamma_k{}^k)^2\bigg)=0.
\eea
On integrating this expression over ${\bcc{\Sigma}}$, and using $\bcn_i {\bcc h}^i=0$,
implies (under the additional assumption that the near-horizon spatial cross section has no boundary) that $\delta \gamma_k{}^k$ is constant.

To analyse the remaining transverse moduli, we linearize the field equations in terms of the moduli
$\delta \Delta, \delta h$, $\delta \gamma$ and $W$, making use of the conditions imposed by supersymmetry which we have previously obtained. We first use the $-i$ and $+-$ components of the Einstein equations to fix the moduli $\delta h$ and $\delta \Delta$ in terms
of $W$ and $\delta \gamma$ as:
\bea
\label{fix1}
\delta h_i =  -{1 \over 2} {\bcc \nabla}_j \delta \gamma_i{}^j
-{1 \over 4}
(\delta \gamma_k{}^k) {\bcc h}_i +{1 \over 2} {\bcc h^j} \delta \gamma_{ij}
+{1 \over 2} \eta {\bcc \Delta}^{1 \over 2} W_i -{1 \over 2} W_j
\epsilon_i{}^{jk} {\bcc h}_k
\eea
\bea
\label{fix2}
\delta \Delta &=& {\bcc \nabla}_i \delta h^i - {\bcc h}^i \delta h_i
- ({\bcc \nabla}_i \delta \gamma^{ij}) {\bcc h}_j
+{1 \over 4} \delta \gamma_k{}^k {\bcc h}_i {\bcc h}^i
\nonumber \\
&-&{1 \over 2} {\bcc \Delta} \delta \gamma_k{}^k
+{1 \over 4} {\bcc h}^i {\bcc h}^j \delta \gamma_{ij}
-{1 \over 2} \eta {\bcc \Delta}^{1 \over 2} {\bcc h}^i W_i
-{1 \over 6} \eta W^i {\bcc \nabla}_i {\bcc \Delta}^{1 \over 2}
-{1 \over 2} \delta \gamma^{ij} {\bcc \nabla}_i {\bcc h}_j.
\eea
We remark that as a consequence of the analysis in \cite{harvey},
the last two terms in ({\ref{fix2}}) vanish, because
\bea
\label{zeroth}
\bcc \Delta = const., \qquad {\bcc \nabla}_{(i} {\bcc h}_{j)}=0.
\eea

Having fixed these moduli, we shall construct elliptic systems of
PDEs constraining $W$ and $\delta \gamma$. To begin, consider the $W$ moduli.
We consider the Bianchi identity
({\ref{Bianchi2}}), which when linearized implies
\bea
{\bcc \nabla}_i W_j - {\bcc \nabla_j} W_i = \bigg({\bcc h} \wedge W
+{\cal{L}}_{\partial \over \partial r} \star_3 Y \bigg)_{ij}.
\eea
On taking the divergence, we then obtain the condition
\bea
{\bcc \nabla}^2 W_j = - {\bcc R}_{ij} W^i
+ {\bcc \nabla_j} \bigg({\bcc \nabla_i} W^i \bigg)
+ {\bcc \nabla}^i \bigg({\bcc h} \wedge W
+{\cal{L}}_{\partial \over \partial r} \star_3 Y \bigg)_{ij}
\eea
and the term ${\bcc \nabla_i} W^i$ is given by linearizing
({\ref{gauge1}}), as
\bea
\label{wdev}
{\bcc \nabla_i} W^i = 3 {\bcc h}^i W_i +{3 \over 2} \eta {\bcc \Delta}^{1 \over 2} \delta \gamma_i{}^i + 6 \eta \delta \big(\Delta^{1 \over 2})
\eea
and hence
\bea
\label{wellip}
{\bcc \nabla}^2 W_j &=& - {\bcc R}_{ij} W^i
+ {\bcc \nabla}^i \bigg({\bcc h} \wedge W \bigg)_{ij}
+ {\bcc \nabla}_j \bigg( 3 {\bcc h}^i W_i +{3 \over 2} \eta {\bcc \Delta}^{1 \over 2} \delta \gamma_i{}^i \bigg)
\nonumber \\
&+& 6 \eta {\bcc \nabla}_j \delta (\Delta^{1 \over 2})
+ {\bcc \nabla}^i \bigg({\cal{L}}_{\partial \over \partial r} \star_3 Y \bigg)_{ij} \bigg|_{r=0}.
\eea
It is clear that the first three terms on the RHS of this expression give
no contribution to the principle symbol of the differential operator
acting on $W$. However, due to the presence of $\delta \Delta$
in the remaining terms, it may appear that the RHS contains terms
of order ${\bcc \nabla}^2 \delta g$ on using ({\ref{fix1}}) and ({\ref{fix2}}). Such terms arise in the combination
\bea
6 \eta {\hat{d}} \delta \Delta^{1 \over 2
}- \star_3 ({\hat d} \delta Y).
\eea
However, on making use of ({\ref{ident3}}), it follows that the
$\delta \Delta$ contribution to this expression vanishing, and hence in
({\ref{wellip}}) the RHS depends on the moduli linearly in
$W, {\bcc \nabla} W, \delta \gamma, {\bcc \nabla} \delta \gamma$.

Next, we consider the $ij$ components of the Einstein equations, which imply
\bea
R_{ij} = {3 \over 2} \big(H_{i+} H_{j-}+H_{j+}H_{i-}+H_{i \ell} H_j{}^\ell\big) -{1 \over 4} \delta_{ij} \big(-2 (H_{+-})^2+4 H_{- \ell} H_+{}^\ell
+H_{\ell_1 \ell_2} H^{\ell_1 \ell_2} \big)
\eea
On making use of ({\ref{ident1}}), all of the terms quadratic
in $H$ on the RHS of this expression depend linearly in
$W, {\bcc \nabla} W, \delta \gamma, {\bcc \nabla} \delta \gamma$, with
the exception of the $(H_{+-})^2$ term, which gives rise to a $\delta \Delta$ term. Taking this into account, and making use of ({\ref{fix1}}) and
({\ref{fix2}}) we find
\bea
\label{einsimp1}
 \bcn^2 \delta \gamma_{ij}  -\delta_{ij} \bcn_k \bcn_\ell \delta \gamma^{\ell k} 
-(\bcn_\ell \bcn_j-\bcn_j \bcn_\ell) \delta \gamma^\ell{}_i
- (\bcn_\ell \bcn_i-\bcn_i \bcn_\ell) \delta \gamma^\ell{}_j = {\cal{A}}_{ij}
\eea
where ${\cal{A}}_{ij}$ depends linearly on $W, {\bcc \nabla} W, \delta \gamma, {\bcc \nabla} \delta \gamma$. This expression can be simplified by first noting that the terms on the second line of the LHS can be rewritten in terms
of ${\bcc R}$ curvature terms, and hence incorporated into the algebraic term on the RHS, i.e.
\bea
\label{einsimp2}
 \bcn^2 \delta \gamma_{ij}  -\delta_{ij}  \bcn_k \bcn_\ell \delta \gamma^{\ell k}  = {\cal{B}}_{ij}
\eea
where ${\cal{B}}_{ij}$ depends linearly on $W, {\bcc \nabla} W, \delta \gamma, {\bcc \nabla} \delta \gamma$. On taking the trace of ({\ref{einsimp2}}) we
also find
\bea
\bcn_k \bcn_\ell \delta \gamma^{\ell k} = -{1 \over 3} {\cal{B}}_i{}^i
\eea
and so the second term on the LHS of ({\ref{einsimp2}}) can be eliminated in favour of ${\cal{B}}_i{}^i$, to give
\bea
\label{einsimp3}
\bcn^2 \delta \gamma_{ij} = {\cal{C}}_{ij}
\eea
where ${\cal{C}}_{ij}$ depends linearly on $W, {\bcc \nabla} W, \delta \gamma, {\bcc \nabla} \delta \gamma$.

The condition ({\ref{einsimp3}}) is an elliptic constraint
on the traceless part of $\delta \gamma_{ij}$. So, we have proven that
there exists a gauge in which the system of
PDEs ({\ref{einsimp3}}), ({\ref{trelliptic}}) together with ({\ref{wellip}})  constitute
an elliptic set of PDEs which constrain the moduli $W$, $\delta \gamma_k{}^k$
and the traceless part of $\delta \gamma_{ij}$.

\koniec

\setcounter{section}{0}
\setcounter{subsection}{0}

\appendix{Spinorial Geometry Conventions}

The space of Dirac spinors consists of the space of complexified forms on $\bR^2$,
which has basis $\{ 1, e_1, e_2, e_{12}=e_1 \wedge e_2 \}$.
We define the action of the Clifford algebra generators on this space via
\begin{eqnarray}
\gamma_i = -e_i \wedge  - i_{e_i}, \qquad \gamma_{i+2}= i \big(-e_i \wedge + i_{e_i} \big) \quad i=1,2
\end{eqnarray}
and set
\begin{eqnarray}
\gamma_0 = i \gamma_{1234}
\end{eqnarray}
which acts as
\begin{eqnarray}
\gamma_0 1 = i 1 , \quad \gamma_0 e_{12} = i e_{12}, \quad \gamma_0 e_i = -i e_i \ .
\end{eqnarray}
We then define generators adapted to the frame ({\ref{frame}}) as
\begin{eqnarray}
\label{fbasis}
\Gamma_\pm = {1 \over \sqrt{2}} (\gamma_3 \pm \gamma_0),
\quad \Gamma_1 = \gamma_1, \quad \Gamma_2 = \sqrt{2} e_2 \wedge,
\qquad \Gamma_{{\bar{2}}} = \sqrt{2} i_{e_2}
\end{eqnarray}
where we take a basis $\{ {{\bf{e}}}^1, {{\bf{e}}}^2, {{\bf{e}}}^{{\bar{2}}} \}$ for ${\Sigma}$ such that ${{\bf{e}}}^{{\bar{2}}}= ({{\bf{e}}}^2)^*$ and
the metric on $\Sigma$ is
\begin{eqnarray}
\gamma = ({{\bf{e}}}^1)^2 +2 {{\bf{e}}}^2 {{\bf{e}}}^{{\bar{2}}} \ .
\end{eqnarray}
With these conventions, the space of positive chirality spinors is spanned by
$\{ 1-e_1, e_2+e_{12} \}$, and the space of negative chirality spinors is spanned by
$\{ 1+e_1, e_2-e_{12} \}$ and we remark that $Spin(3)$, with generators
$i \Gamma_{2 {{\bar{2}}}}, \Gamma_1(\Gamma_2+\Gamma_{{\bar{2}}}),
i  \Gamma_1(\Gamma_2-\Gamma_{{\bar{2}}})$ form a representation of $SU(2)$ acting on
$\{1-e_1, e_2+e_{12} \}$.

A $Spin(4,1)$ invariant inner product $\beta$ on the space of spinors is then given by
\begin{eqnarray}
\label{prod}
\beta (\epsilon_1, \epsilon_2) = \langle \gamma_0 \epsilon_1, \epsilon_2 \rangle = {1 \over \sqrt{2}} \langle (\Gamma_+ - \Gamma_-) \epsilon_1, \epsilon_2 \rangle
\end{eqnarray}
where $\langle , \rangle$ denotes the canonical inner product on $\bC^4$ equipped with basis
$\{1, e_1, e_2, e_{12} \}$.

The charge conjugation operator $C$ is defined by
\bea
C.1 = -e_{12} \ , \quad C.e_{12}=1 \ , \quad C. e_i = \epsilon_i{}^j e_j
\eea
and satisfies
\bea
C* \Gamma_\mu + \Gamma_\mu C* =0.
\eea
With respect to the {\it{real}} frame ({\ref{frame}}),
\bea
\Gamma_i^\dagger = \Gamma_i, \quad \Gamma_+^\dagger = \Gamma_-, \quad \Gamma_-^\dagger = \Gamma_+.
\eea
We also note the following useful identities:
\begin{eqnarray}
\Gamma_{ij} \epsilon_\pm = \mp i \epsilon_{ij}{}^k \Gamma_k \epsilon_\pm,
\quad \Gamma_{ijk} \epsilon_\pm = \mp i \epsilon_{ijk} \epsilon_\pm.
\end{eqnarray}
The relationship between the 5-dimensional volume form
$\epsilon_5$ and the volume form $\epsilon_{\Sigma}$ of ${\Sigma}$ is
\bea
\epsilon_5 = \bbe^+ \wedge \bbe^- \wedge \epsilon_{{\Sigma}}\;.
\eea

\appendix{Ricci Tensor}

The components of the Ricci tensor in this basis are:
\bea
{\mathcal R}_{++} &=& r^2 \bigg(-{3 \over 2} h^i \hn_i \Delta-{1 \over 2}
\Delta \hn_i h^i +{1 \over 2} \hn_i \hn^i \Delta +
\Delta h_i h^i +{1 \over 4} ({\hat{d}}h)_{ij}
({\hat{d}}h)^{ij}\bigg)
\nonumber \\
&+& r^3 \bigg({1 \over 2} \dh^i \tn_i \Delta- h^i
\tn_i {\dot{\Delta}} +{1 \over 2} h^i \tn^j \Delta
{\dot{g}}_{ij}
-{1 \over 4} \dot{g}_k{}^k h^i \tn_i \Delta+2 {\dot{\Delta}}
h_i h^i
\nonumber \\
 &-&\Delta h_i \dh^i
- h^i \dh^j ({\hat{d}}h)_{ij}
-{1 \over 4} \Delta h^i h^j {\dot{g}}_{ij}
+{1 \over 4} \Delta h_i h^i {\dot{g}}_k{}^k
+{1 \over 2} \Delta \tn_i \dh^i -{1 \over 2}
{\dot{\Delta}} \tn_i h^i \bigg)
\nonumber \\
&+& r^4 \bigg(-{1 \over 2} \Delta h^i \ddh_i
+{1 \over 2} \Delta h^i \dh^j {\dot{g}}_{ij}
- {1 \over 2} {\dot{\Delta}} h^i h^j {\dot{g}}_{ij}
-{1 \over 8} \Delta^2 \big( \partial_r {\dot{g}}_k{}^k
+{1 \over 2} {\dot{g}}_{ij} {\dot{g}}^{ij} \big)
\nonumber \\
&+&{1 \over 2} {\ddot{\Delta}} h_i h^i
+{1 \over 4} {\dot{\Delta}} h_i h^i {\dot{g}}_k{}^k
-{1 \over 4} \Delta h_i \dh^i {\dot{g}}_k{}^k
+{1 \over 2} h_i h^i \dh_j \dh^j
-{1 \over 2} h^i \dh_i h^j \dh_j \bigg)
\eea
\bea
{\mathcal R}_{+-} &=& {1 \over 2} \tn_i h^i
-{1 \over 2} h_i h^i - \Delta
\nonumber \\
&+& r \bigg( {1 \over 2} \tn_i \dh^i
-{1 \over 4} {\dot{g}}_k{}^k h_j h^j
-2 {\dot{\Delta}} - {1 \over 2} \Delta {\dot{g}}_k{}^k
+{1 \over 2} h^i h^j {\dot{g}}_{ij} -2 h^i \dh_i\bigg)
\nonumber \\
&+& r^2 \bigg( -{1 \over 2} h^i \ddh_i
+{1 \over 2} h^i \dh^j {\dot{g}}_{ij}
-{1 \over 2} \dh_i \dh^i -{1 \over 4} {\dot{g}}_k{}^k h_j h^j
\nonumber \\
&-&{1 \over 2}{\ddot{\Delta}} -{1 \over 4} {\dot{\Delta}} {\dot{g}}_k{}^k
-{1 \over 4} \Delta \big(\partial_r {\dot{g}}_k{}^k+{1 \over 2} {\dot{g}}_{ij}
{\dot{g}}^{ij} \big) \bigg)
\eea
\bea
{\mathcal R}_{--} = -{1 \over 2} \big(\partial_r {\dot{g}}_k{}^k
+{1 \over 2} {\dot{g}}_{ij} {\dot{g}}^{ij}\big)
\eea
\bea
{\mathcal R}_{-i} &=& \dh_i -{1 \over 2} \tn_i {\dot{g}}_k{}^k
+{1 \over 4} h_i {\dot{g}}_k{}^k
-{1 \over 2} h^j {\dot{g}}_{ij} +{1 \over 2} \tn_j
{\dot{g}}_i{}^j
\nonumber \\
&+& r \bigg( {1 \over 2} \ddh_i -{1 \over 2} h^j
{\ddot{g}}_{ij} + h_i \big({1 \over 2}
\partial_r  {\dot{g}}_k{}^k +{1 \over 4} {\dot{g}}_{jk}
{\dot{g}}^{jk} \big)
\nonumber \\
&+&{1 \over 4} {\dot{g}}_k{}^k \dh_i -{1 \over 2}
\dh^j {\dot{g}}_{ij} -{1 \over 4} {\dot{g}}_k{}^k
h^j {\dot{g}}_{ij} \bigg)
\eea
\bea
{\mathcal R}_{+i} &=& r \bigg({1 \over 2} \tn_j ({\hat{d}}h)_i{}^j
-h^j ({\hat{d}}h)_{ij}+ \Delta h_i  - \tn_i \Delta \bigg)
\nonumber \\
&+& r^2 \bigg(-{1 \over 2} \Delta \dh_i -{1 \over 2} h_i
\tn_j \dh^j +{1 \over 2} \dh_i \tn_j h^j
+ h^j \tn_j \dh_i -{1 \over 2} \tn_i (h_j \dh^j) +{1 \over 2} {\dot{g}}_i{}^j \tn_j \Delta
\nonumber \\
&+&{1 \over 4} \Delta \tn_j {\dot{g}}_i{}^j
+{1 \over 2} {\dot{g}}_i{}^k ({\hat{d}}h)_k{}^j h_j
+{1 \over 2} ({\hat{d}}h)_{ik} h^j {\dot{g}}_j{}^k
+{3 \over 2} h_j h_i \dh^j -{3 \over 2} h_j h^j \dh_i
-{3 \over 4} \Delta {\dot{g}}_i{}^j h_j
\nonumber \\
&-&{1 \over 2} \tn_i {\dot{\Delta}}
-{1 \over 4} \tn_i (\Delta {\dot{g}}_k{}^k)
+2 {\dot{\Delta}} h_i +{3 \over 8} \Delta h_i
{\dot{g}}_k{}^k -{1 \over 4} h^j ({\hat{d}}h)_{ij}
{\dot{g}}_k{}^k \bigg)
\nonumber \\
&+& r^3 \bigg(-{1 \over 4} \Delta \ddh_i
+{1 \over 2} h_i h^j \ddh_j -{1 \over 2} h_j h^j \ddh_i
+{1 \over 2} h_i \dh_j \dh^j -{1 \over 2} h_j \dh^j \dh_i
+{1 \over 4} \Delta {\dot{g}}_i{}^j \dh_j
\nonumber \\
&+&{1 \over 2}{\dot{g}}_i{}^k
(h_j h^j \dh_k - h_j \dh^j h_k)
+{1 \over 2} (\dh_i h^j h^k - h_i h^j \dh^k) {\dot{g}}_{jk}
-{1 \over 4} {\dot{\Delta}} {\dot{g}}_i{}^j h_j
\nonumber \\
&+& {1 \over 2} {\ddot{\Delta}} h_i +{1 \over 4} h_i {\dot{\Delta}} {\dot{g}}_k{}^k
+{1 \over 2} h_i \Delta \big({1 \over 2} \partial_r
{\dot{g}}_k{}^k +{1 \over 4}{\dot{g}}_{jk} {\dot{g}}^{jk} \big)+{1 \over 4} \dh_i {\dot{\Delta}}-{1 \over 4}
{\dot{\Delta}} {\dot{g}}_i{}^j h_j -{1 \over 8} \Delta \dh_i
{\dot{g}}_k{}^k
\nonumber \\
&+&{1 \over 2} {\dot{g}}_k{}^k h_j \big({1 \over 2} h_i
\dh^j -{1 \over 2} h^j \dh_i -{1 \over 4} \Delta
{\dot{g}}_i{}^j \big)-{1 \over 4} \Delta
{\ddot{g}}_{ij} h^j \bigg)
\eea
\bea
{\mathcal R}_{ij} &=& {\hat{{\mathcal R}}}_{ij}
+\tn_{(i} h_{j)} -{1 \over 2} h_i h_j
\nonumber \\
&+& r \bigg(\tn_{(i} \dh_{j)}-3h_{(i} \dh_{j)}
+\big(-\Delta+{1 \over 2}\tn_k h^k -h_k h^k\big)
{\dot{g}}_{ij}
-{\dot{g}}_{(i}{}^k \tn_{|k|} h_{j)}
\nonumber \\
&-& h^k \tn_{(i} {\dot{g}}_{j) k}
+h^k \tn_k {\dot{g}}_{ij}
-h_{(i} \tn^k {\dot{g}}_{j) k} +2 h_k h_{(i} {\dot{g}}_{j)}{}^k+ h_{(i} \tn_{j)} {\dot{g}}_k{}^k
\nonumber \\
&+&{1 \over 2}{\dot{g}}_k{}^k \big(\tn_{(i} h_{j)}-h_i h_j \big) \bigg)
\nonumber \\
&+& r^2 \bigg( -{1 \over 2}(\Delta+h_k h^k)
\big({\ddot{g}}_{ij} - {\dot{g}}_i{}^n {\dot{g}}_{nj}\big)
+{1 \over 2} h_i h_n \big({\ddot{g}}_j{}^n - {\dot{g}}_j{}^k {\dot{g}}_k{}^n \big)
+{1 \over 2} h_j h_n \big({\ddot{g}}_i{}^n - {\dot{g}}_i{}^k {\dot{g}}_k{}^n \big)
\nonumber \\
&-&{1 \over 2}{\dot{\Delta}} {\dot{g}}_{ij}
-{1 \over 2} h_i \ddh_j -{1 \over 2} h_j \ddh_i
-{1 \over 2} h_i h_j \big( \partial_r {\dot{g}}_k{}^k
+{1 \over 2} {\dot{g}}_{nk} {\dot{g}}^{nk} \big)
-{1 \over 2} \dh_i \dh_j -{1 \over 4} \Delta {\dot{g}}_k{}^k {\dot{g}}_{ij}
\nonumber \\
&+&{1 \over 2} {\dot{g}}_i{}^k (h_j \dh_k + h_k \dh_j)
+ {1 \over 2} {\dot{g}}_j{}^k (h_i \dh_k + h_k \dh_i)
-h_k \dh^k {\dot{g}}_{ij} -{1 \over 4} {\dot{g}}_k{}^k
(h_i \dh_j + h_j \dh_i)
\nonumber \\
&-&{1 \over 4} h_n h^n {\dot{g}}_k{}^k {\dot{g}}_{ij}
+{1 \over 4} {\dot{g}}_k{}^k h_i h_n {\dot{g}}_j{}^n
+{1 \over 4} {\dot{g}}_k{}^k h_j h_n {\dot{g}}_i{}^n
+{1 \over 2} h^n h^k {\dot{g}}_{nk} {\dot{g}}_{ij}
-{1 \over 2} h^n h^k {\dot{g}}_{in} {\dot{g}}_{jk} \bigg)
\eea
Here ${\hat{{\mathcal R}}}_{ij}$ denotes the Ricci tensor of ${\Sigma}$, and $\ \dot{} \ $ denotes the Lie derivative
with respect to ${\partial \over \partial r}$,
so
\bea
\dh = {\cal{L}}_{\partial \over \partial r} h,
\qquad \de^i = {\cal{L}}_{\partial \over \partial r} \bbe^i , \qquad
\ddh = {\cal{L}}_{\partial \over \partial r} \dh , \qquad
 {\rm etc}.
\eea
and
\bea
\partial_r (h_i) = \dh_i - (\de^j)_i h_j,
\qquad \partial_r (\dh_i) = \ddh_i - (\de^j)_i \dh_j \ .
\eea

\appendix{Analysis of Gravitino Equation}

In this Appendix we present the analysis of the gravitino Killing spinor equation.
We first list the components of the spin connection, and then investigate the
gravitino equation

\subsection{The spin connection}

With respect to the frame ({\ref{frame}}) we have
\bea
d \bbe^+ &=& 0,
\nonumber \\
d \bbe^- &=&
(\bbe^- -{1 \over 2} r^2 \Delta \bbe^+) \wedge h + r {\hat{d}} h
+r \Delta \bbe^+ \wedge \bbe^- -{1 \over 2} r^2 {\hat{d}} \Delta \wedge \bbe^+
\nonumber \\
&+&{1 \over 2} r^2 {\dot \Delta}  \bbe^+ \wedge \bbe^- +{1 \over 2} r^3 {\dot \Delta} h \wedge \bbe^+
+r (\bbe^- -rh +{1 \over 2} r^2 \Delta \bbe^+) \wedge {\dot{h}},
\nonumber \\
d \bbe^i &=& (\bbe^- -rh +{1 \over 2} r^2 \Delta \bbe^+) \wedge {\dot{\bf{e}}}^i  +{\hat{d}} \bbe^i.
\eea

Also, if $g$ is any function, then the relationship between frame and co-ordinate indices is:
\bea
\partial_+ g &=& \partial_u g +{1 \over 2} r^2 \Delta {\dot{g}}
\nn
\partial_- g &=& {\dot{g}}
\nn
\partial_i g &=& {\hat{\partial}}_i g -r {\dot{g}} h_i
\eea
where ${\hat{\partial}}_i = e^j{}_i \partial_j$. It follows that the components of the spin connection are given by
\bea
\omega_{+,+-} &=& -r \Delta -{1 \over 2} r^2 {\dot{\Delta}}
\nonumber \\
\omega_{+,+i} &=& {1 \over 2} r^2 \Delta h_i -{1 \over 2} r^2 \hp_i \Delta
+{1 \over 2} r^3 {\dot{\Delta}} h_i -{1 \over 2} r^3 \Delta \dot{h}_i
\nonumber \\
\omega_{+,-i} &=& -{1 \over 2} h_i -{1 \over 2}r {\dot{h}}_i
\nonumber \\
\omega_{+,ij} &=& -{1 \over 2} r ({\hat{d}} h)_{ij} +{1 \over 2} r^2 (h_i {\dot{h}}_j - h_j {\dot{h}}_i)
-{1 \over 4} r^2 \Delta (\dot \bbe^i)_j +{1 \over 4} r^2 \Delta (\dot \bbe^j)_i
\nonumber \\
\omega_{-,+-} &=&0
\nonumber \\
\omega_{-,+i} &=& -{1 \over 2} h_i -{1 \over 2} r {\dot h}_i
\nonumber \\
\omega_{-,-i} &=& 0
\nonumber \\
\omega_{-,ij} &=& -{1 \over 2} (\dot \bbe^i)_j +{1 \over 2} (\dot \bbe^j)_i
\nonumber \\
\omega_{i,+-} &=& {1 \over 2} h_i +{1 \over 2} r {\dot h}_i
\nonumber \\
\omega_{i,+j} &=& -{1 \over 4} r^2 \Delta \big( (\dot \bbe^i)_j + (\dot \bbe^j)_i \big)
-{1 \over 2} r ({\hat{d}} h)_{ij} +{1 \over 2} r^2 (h_i {\dot h}_j - h_j {\dot h}_i)
\nonumber \\
\omega_{i,-j} &=& -{1 \over 2} \big( (\dot \bbe^i)_j +(\dot \bbe^j)_i \big)
\nonumber \\
\omega_{i,jk} &=& \Omega_{i,jk} +{1 \over 2} r h_i \big( (\dot \bbe^j)_k -(\dot \bbe^k)_j \big)
+{1 \over 2} r h_j  \big((\dot \bbe^i)_k + (\dot \bbe^k)_i \big)
-{1 \over 2} r h_k \big( (\dot \bbe^i)_j + (\dot \bbe^j)_i \big)
\eea
where $\Omega_{i,jk}$ is the spin connection of ${\Sigma}$ with basis $\bbe^i$
(restricting to constant $r$).

\subsection{Analysis of the KSE}

Next, we consider the gravitino KSE ({\ref{grav}}).
We shall first analyse these equations acting on a $u$-independent spinor $\epsilon=\epsilon_+ + \epsilon_-$,
and then apply the conditions ({\ref{bl1}}), ({\ref{bl2}}) and ({\ref{bl3}}) imposed
by the bi-linear matching.
We begin by analysing the $\alpha=-$ and $\alpha=+$ components of
({\ref{grav}}).

From the $\alpha=-$ component of ({\ref{grav}}) we
obtain the conditions
\bea
\label{rd1}
\partial_r \epsilon_+ = \bigg({1 \over 4} ({\dot{\bbe}}^i)_j \Gamma_i{}^j -{3i \over 4} H_{-i} \Gamma^i \bigg) \epsilon_+
\eea
and
\bea
\label{rd2}
\partial_r \epsilon_- &=& \Gamma_- \bigg({1 \over 4}
(h_i+ r {\dot h}_i) \Gamma^i
+{i \over 2} H_{+-} +{i \over 8} H_{ij} \Gamma^{ij} \bigg) \epsilon_+
\nonumber \\
&+&\bigg({1 \over 4} ({\dot{\bbe}}^i)_j \Gamma_i{}^j
-{i \over 4} H_{-i} \Gamma^i \bigg) \epsilon_-.
\eea
Furthermore, on substituting ({\ref{rd1}}) and ({\ref{rd2}})
into the $\alpha=+$ component of ({\ref{grav}}), and
using the condition $\partial_u \epsilon=0$ which we have
obtained previously, we find the following algebraic
conditions
\bea
\label{algc1}
\bigg(-{3i \over 8} r^2 \Delta H_{-i} \Gamma^i
+{1 \over 2}(r \Delta +{1 \over 2}r^2 {\dot \Delta})
+ \big(-{1 \over 8}r ({\hat{d}}h)_{ij} +{1 \over 4}r^2 h_i
{\dot h}_j \big) \Gamma^{ij}
+{i \over 4} H_{+i} \Gamma^i  \bigg) \epsilon_+
\nonumber \\
+ \Gamma_+ \bigg(-{1 \over 4} (h_i + r {\dot h}_i) \Gamma^i -{i \over 8} H_{ij}\Gamma^{ij} +{i \over 2} H_{+-}\bigg) \epsilon_-=0
\eea
and
\bea
\label{algc2}
\Gamma_- \bigg({i \over 4} r^2 \Delta H_{+-}
+{i \over 16} r^2 \Delta H_{ij} \Gamma^{ij}
+\big({3 \over 8}r^2 \Delta h_i -{1 \over 4} r^2 \hn_i \Delta
+{1 \over 4} r^3 {\dot \Delta} h_i
-{1 \over 8} r^3 \Delta {\dot h}_i \big) \Gamma^i \bigg) \epsilon_+
\nonumber \\
+ \bigg(-{i \over 8}r^2 \Delta H_{-i} \Gamma^i
-{1 \over 2}(r \Delta +{1 \over 2}r^2 {\dot \Delta})
+ \big(-{1 \over 8}r ({\hat{d}} h)_{ij} +{1 \over 4} r^2
h_i {\dot h}_j \big) \Gamma^{ij}
+{3i \over 4} H_{+i}\Gamma^i  \bigg)
\epsilon_-=0.
\eea

This exhausts the content of the $\alpha=+$ and $\alpha=-$ components
of ({\ref{grav}}).
The $\alpha=i$ component of ({\ref{grav}}) is equivalent to
\bea
\label{sp1}
\hn_i \epsilon_+ + \bigg({3i \over 4} r h_i H_{-j} \Gamma^j
-{1 \over 4}(h_i+r {\dot h}_i)
+{1 \over 4} r h_j {\dot \gamma}_{ik}\Gamma^{jk}
+{i \over 4} \Gamma_i H_{+-}
-{i \over 8} \Gamma_i{}^{jk}  H_{jk} +{i \over 2} H_{ij} \Gamma^j
 \bigg) \epsilon_+
\nonumber \\
+ \bigg({1 \over 4} {\dot \gamma}_{ij} \Gamma^j +{i \over 4} H_{-j} \Gamma_i{}^j
-{i \over 2} H_{-i} \bigg) \Gamma_+ \epsilon_- =0
\eea
\bea
\label{sp2}
\hn_i \epsilon_- + \bigg({i \over 4} r h_i H_{-j} \Gamma^j
+{1 \over 4}(h_i+r{\dot h}_i)+{1 \over 4} r h_j {\dot \gamma}_{ik}
\Gamma^{jk}
-{i \over 4} \Gamma_i H_{+-}
-{i \over 8} \Gamma_i{}^{jk} H_{jk} +{i \over 2} H_{ij} \Gamma^j \bigg) \epsilon_-
\nonumber \\
+ \bigg({1 \over 4} r h_i h_j \Gamma^j -{i \over 2} r h_i H_{+-}-{i \over 8} r h_i
H_{jk} \Gamma^{jk}
+ \big({1 \over 8} r^2 \Delta {\dot \gamma}_{ij} +{1 \over 4} r
({\hat{d}} h)_{ij}+{1 \over 4} r^2 {\dot h}_i h_j \big) \Gamma^j
\nonumber \\
+{i \over 4} H_{+j} \Gamma_i{}^j -{i \over 2} H_{+i} \bigg) \Gamma_- \epsilon_+ =0
\nonumber \\
\eea

where
\bea
{\dot \gamma}_{ij} =\bbe^m{}_i \bbe^n{}_j \delta_{k \ell}
\bigg( {\dot \bbe^k}{}_m \bbe^\ell{}_n + \bbe^k{}_m {\dot \bbe^\ell}{}_n \bigg).
\eea

\subsection{$u$-dependence of the Spinor and Bi-linear Matching}

To proceed, note first that if $\epsilon$ is a Killing spinor then so is $C* \epsilon$, where
$C*$ denotes the charge conjugation operator. Also, $\epsilon$ and $C* \epsilon$ are
linearly independent (over $\bC$). We shall assume that the bulk black hole solution
is half-supersymmetric.

As all the bosonic fields, and the frame, are $u$-independent, it follows that if $\epsilon$ is a Killing spinor then so is
$\partial_u \epsilon$. This implies that there exist constants $k_1, k_2 \in \bC$ such that
\bea
\partial_u \epsilon = k_1 \epsilon + k_2 C* \epsilon.
\eea
Now consider the Killing spinor
\bea
{\tilde{\epsilon}}  = \alpha \epsilon + \beta C* \epsilon
\eea
for constant $\alpha, \beta \in \bC$. By choosing  $\alpha, \beta$
appropriately (not both zero), it follows that there exists a Killing spinor ${\tilde{\epsilon}}$
such that
\bea
\partial_u \tilde{\epsilon} = k \tilde{\epsilon}
\eea
for constant $k \in \bC$, and hence
\bea
{\tilde{\epsilon}} = e^{k u} \phi
\eea
where $\partial_u \phi =0$.
If the solution is exactly half-supersymmetric, then any Killing spinor $\epsilon$ can be written as a linear combination of ${\tilde{\epsilon}}$
and $C* {\tilde{\epsilon}}$, it follows that
\bea
\label{udep2}
\epsilon = \ell_1 e^{ku} \phi +\ell_2 e^{{\bar{k}} u} C* \phi
\eea
for complex constants $\ell_1, \ell_2$. We shall require that
the spinor has a well-defined near-horizon limit, which
implies that $k=0$, and hence the Killing spinor
$\epsilon$ is $u$-independent.

Now, we shall assume that there exists a Killing spinor $\epsilon$
such that the $Spin(4,1)$-invariant 1-form Killing spinor bi-linear $Z$,
where
\bea
\label{kvc1}
Z_\alpha= \langle (\Gamma_+ - \Gamma_-) \epsilon, \Gamma_\alpha \epsilon \rangle
\eea
is proportional to the 1-form dual to the black hole Killing vector
\bea
\label{kvc2}
V = {\partial \over \partial u}.
\eea
We shall identify $V$ with $Z$, and set, without loss of generality
\bea
Z = -2V = r^2 \Delta \bbe^+ -2 \bbe^-.
\eea

In order to impose the bi-linear matching condition $Z=-2V$,
we decompose the spinor $\epsilon$ as
\bea
\epsilon = \epsilon_+ + \epsilon_-, \qquad \Gamma_\pm \epsilon_\pm =0 \ .
\eea
The condition $Z_-=-2V_-$ implies
\bea
\label{bl1}
\parallel \epsilon_+ \parallel^2 =1
\eea
and the condition  $Z_+ =-2 V_+$ implies
\bea
\label{bl2}
\parallel \epsilon_- \parallel^2 = {1 \over 2} r^2 \Delta
\eea
We also require $Z_i=0$, or equivalently
\bea
\label{bl3a}
{\rm Re} \ \bigg( \langle \epsilon_+, \Gamma_i \Gamma_+ \epsilon_- \rangle \bigg) =0.
\eea
We shall use spinorial geometry methods to analyse this condition.
First, note that we can apply a $Spin(3)$ gauge transformation as described in the
Appendix A, to set
\bea
\epsilon_+ = f (1-e_1)
\eea
where $f \in \bR$. Note that ({\ref{bl1}}) implies that $2f^2=1$.
In addition, we set
\bea
\epsilon_- = p (1+e_1)+q (e_2-e_{12})
\eea
for $p,q \in \bC$. So ({\ref{bl3a}}) can be rewritten as
\bea
{\rm Im} \ \bigg( \langle \Gamma_i (1+e_1),  p (1+e_1)+q (e_2-e_{12}) \rangle \bigg) =0.
\eea
It is straightforward to note that these conditions imply that
\bea
{\rm Im} \ ( p )=0, \qquad q =0.
\eea
Hence
\bea
\epsilon_- = h (1+e_1)
\eea
for $h \in \bR$. The condition ({\ref{bl2}}) implies that $2h^2 = {1 \over 2} r^2 \Delta$.
In particular, note that $\Delta \geq 0$.
These conditions can be rewritten as
\bea
\epsilon_- = {i \over \sqrt{2}} {h \over f} \Gamma_- \epsilon_+
\eea
and on using ({\ref{bl1}}) and ({\ref{bl2}}), this can be further rewritten as
\bea
\label{bl3}
\epsilon_- = {i \over 2} \eta r \Delta^{1 \over 2} \Gamma_- \epsilon_+
\eea
where $\eta^2=1$. We then compute the gauge-invariant scalar bi-linear,
to obtain
\bea
\beta(\epsilon,\epsilon)= -\sqrt{2} i \eta r \sqrt{\Delta} \ .
\eea
We require that all spinor bilinears are analytic functions
of $r$, and so $r \sqrt{\Delta}$ is analytic in $r$.

\subsection{Bi-linear Matching: Further Simplification of the KSE}

Next we substitute the bi-linear matching conditions ({\ref{bl1}}), ({\ref{bl2}}) and ({\ref{bl3}}) into
the KSE conditions ({\ref{rd2}}), ({\ref{algc1}}), ({\ref{algc2}}), ({\ref{sp1}}), ({\ref{sp2}}).
We rewrite these conditions in terms of conditions solely on $\epsilon_+$.

First, note that on using ({\ref{rd1}}) it follows that  ({\ref{rd2}}), ({\ref{algc1}}), ({\ref{algc2}}) are equivalent to
\bea
\label{alx1}
H_{+-} = \eta (\Delta^{1 \over 2} +{1 \over 2} r
\Delta^{-{1 \over 2}} {\dot \Delta} )
\eea
\bea
\label{alx2}
{1 \over 4}(h_i + r {\dot h}_i)
+{1 \over 8} \epsilon_i{}^{mn} H_{mn} -{1 \over 2}
\eta r \Delta^{1 \over 2} H_{-i} =0
\eea
\bea
\label{alx3}
-{3 \over 8} r^2 \Delta H_{-i} + \big({1 \over 8} r
({\hat{d}} h)_{mn}-{1 \over 4}r^2 h_m {\dot h}_n
\big) \epsilon_i{}^{mn}
+{1 \over 4} H_{+i}
 + \eta r \Delta^{1 \over 2}
({1 \over 4}(h_i +r {\dot h}_i)-{1 \over 8}
\epsilon_i{}^{mn} H_{mn}) =0
\nonumber \\
\eea
\bea
\label{alx4}
{1 \over 16} r^2 \Delta \epsilon_i{}^{mn} H_{mn}
+{3 \over 8} r^2 \Delta h_i -{1 \over 4} r^2 {\hat{\nabla}}_i \Delta +{1 \over 4} r^3 {\dot \Delta} h_i -{1 \over 8} r^3 \Delta {\dot h}_i
\nonumber \\
-{1 \over 16} \eta r^3 \Delta^{3 \over 2} H_{-i}
-{1 \over 2} \eta r \Delta^{1 \over 2}\big({1 \over 8} r
({\hat{d}} h)_{mn}-{1 \over 4}r^2 h_m {\dot h}_n
\big) \epsilon_i{}^{mn} +{3 \over 8} \eta r \Delta^{1 \over 2} H_{+i} =0
\eea
and it is straightforward to show that
the KSE ({\ref{sp2}}) is implied by ({\ref{sp1}})
together with ({\ref{alx1}}), ({\ref{alx2}}), ({\ref{alx3}})
and ({\ref{alx4}}).
The conditions ({\ref{alx1}}), ({\ref{alx2}}), ({\ref{alx3}})
and ({\ref{alx4}}) determine all of the components of $H$, as
\bea
\label{hpm}
H_{+-} = \eta (\Delta^{1 \over 2} +{1 \over 2} r
\Delta^{-{1 \over 2}} {\dot \Delta} )
\eea
\bea
\label{hmi}
H_{-i} &=& {1 \over 3} \eta r^{-1} \Delta^{-{1 \over 2}} h_i
+{2 \over 3} \eta \Delta^{-{1 \over 2}} {\dot h}_i
+{1 \over 6} \eta r^{-1} \Delta^{-{3 \over 2}} {\hat{\nabla}}_i \Delta
\nonumber \\
&-&{1 \over 6} \eta \Delta^{-{3 \over 2}} {\dot \Delta} h_i
+{4 \over 3} r^{-2} \Delta^{-1}
\big({1 \over 8} r
({\hat{d}} h)_{mn}-{1 \over 4}r^2 h_m {\dot h}_n
\big) \epsilon_i{}^{mn}
\eea
\bea
\label{hij}
{1 \over 2} \epsilon_i{}^{mn} H_{mn}
&=& -{1 \over 3} h_i +{1 \over 3} r {\dot h}_i
+{1 \over 3} \Delta^{-1} {\hat{\nabla}}_i \Delta-{1 \over 3}
r \Delta^{-1} {\dot \Delta} h_i
\nonumber \\
&+&{8 \over 3} \eta r^{-1} \Delta^{-{1 \over 2}}
\big({1 \over 8} r
({\hat{d}} h)_{mn}-{1 \over 4}r^2 h_m {\dot h}_n
\big) \epsilon_i{}^{mn}
\eea
\bea
\label{hpi}
H_{+i} &=& -{5 \over 6} \eta r \Delta^{1 \over 2} h_i
+{1 \over 3} \eta r^2 \Delta^{1 \over 2} {\dot h}_i
+{7 \over 12} \eta r \Delta^{-{1 \over 2}} {\hat{\nabla}}_i
\Delta -{7 \over 12} \eta r^2 \Delta^{-{1 \over 2}} {\dot \Delta} h_i
\nonumber \\
&+&{2 \over 3} \big({1 \over 8} r
({\hat{d}} h)_{mn}-{1 \over 4}r^2 h_m {\dot h}_n
\big) \epsilon_i{}^{mn}
\eea
which implies that
\bea
\label{max}
H &=& \eta du \wedge d (r \Delta^{1 \over 2}) +{2 \over 3} \eta d \big(\Delta^{-{1 \over 2}} h \big)
+{1 \over 3} \eta \Delta^{-{3 \over 2}} r^{-1} dr \wedge
d \Delta
\nonumber \\
&+&{1 \over 3} r^{-1}(dr+rh) \wedge \bigg(
\Delta^{-1}\star_3 ({\hat{d}} h -r h \wedge {\dot h})
+ \eta \Delta^{-{1 \over 2}} h -{1 \over 2} \eta
\Delta^{-{3 \over 2}} d \Delta \bigg)
\nonumber \\
&+& {1 \over 3} \star_3 \bigg(-h+r {\dot h}
+ \Delta^{-1} {\hat{d}} \Delta -r \Delta^{-1}
{\dot \Delta} h \bigg) \ ,
\eea
or equivalently
\bea
H= \eta du \wedge d \big(r \Delta^{1 \over 2} \big)
+{1 \over 3} \bigg(\star_3 Y + (dr+rh) \wedge W \bigg)
\eea
where
\bea
W_i = 3H_{-i} \ , \qquad Y_i = {3\over2}\epsilon_i{}^{mn}H_{mn} \ .
\eea

\subsection{Gauge Field Equations}

Here, we briefly summarize some details required for the evaluation of the gauge field equations. In particular, we have

\bea
\star H &=& {1 \over 3} r du \wedge dh -{1 \over 3} du \wedge dr \wedge h - \eta (\Delta^{1 \over 2}
+{1 \over 2} r \Delta^{-{1 \over 2}} {\dot \Delta})
\ {\rm dvol}_{{\cal{S}}}
\nonumber \\
&+& du \wedge (dr+rh) \wedge \bigg({1 \over 3} \Delta^{-1}
d \Delta +{2 \over 3} \eta \Delta^{-{1 \over 2}}
\star_3 ({\hat{d}} h -r h \wedge {\dot h}) \bigg)
\nonumber \\
&-& {2 \over 3} \eta r \Delta^{1 \over 2}
du \wedge \star_3 \bigg(h-r {\dot h}- \Delta^{-1}
{\hat{d}} \Delta +r \Delta^{-1} {\dot \Delta} h\bigg)
\nonumber \\
&-&(dr +rh) \wedge \bigg( \star_3 \big({1 \over 3}\eta r^{-1}
\Delta^{-{1 \over 2}} h +{2 \over 3} \eta \Delta^{-{1 \over 2}}
{\dot h} +{1 \over 6} \eta r^{-1} \Delta^{-{3 \over 2}} {\hat{d}} \Delta -{1 \over 6} \eta \Delta^{-{3 \over 2}}
{\dot \Delta} h \big)
\nonumber \\
&+&{1 \over 3}r^{-1} \Delta^{-1} dh  \bigg) \ ,
\eea
or equivalently
\bea
\star H &=&-du \wedge d(rh) +{2 \over 3} \eta r \Delta^{1 \over 2}du \wedge \bigg(\star_3 Y + (dr+rh) \wedge W \bigg)
\nonumber \\
&-&{1 \over 3} (dr +rh) \wedge \star_3 W - \eta \partial_r \big(r \Delta^{1 \over 2} \big)
{\rm dvol}_{{\Sigma}} \ .
\eea

Then the $(urij)$ component of the gauge equations is equivalent to ({\ref{Bianchi1}}), and the
$(uijk)$ component is equivalent to ({\ref{Bianchi2}}). The only remaining $(rijk)$ component of the gauge equations then reduces to ({\ref{gauge1}}).

\vskip 0.5cm
\noindent{\bf Acknowledgements} \vskip 0.1cm
\noindent  JG is supported by the STFC Consolidated Grant ST/L000490/1. 
The work of WS is supported in part by the National Science
Foundation under grant number PHY-1415659.
MD and JG thank the American University of Beirut for hospitality when some of this work undertaken.
\vskip 0.5cm


\end{document}